\documentclass[conference]{IEEEtran}
\IEEEoverridecommandlockouts
\usepackage{amsmath}
\usepackage{amsfonts}
\usepackage{graphicx}
\usepackage{booktabs}
\usepackage{url}
\usepackage{cite}
\usepackage{acronym} 
\usepackage{tikz}
\usetikzlibrary{positioning,fit,shapes,arrows.meta,calc,matrix}
\usepackage{threeparttable}
\usepackage[hidelinks]{hyperref}

\title{PROTECT-90: A Fault Dataset for Power System Protection%
\thanks{\textcopyright{} 2026 IEEE. Personal use of this material is permitted.  Permission from IEEE must be obtained for all other uses, in any current or future media, including reprinting/republishing this material for advertising or promotional purposes, creating new collective works, for resale or redistribution to servers or lists, or reuse of any copyrighted component of this work in other works.}
}

\author{
\IEEEauthorblockN{
    Julian Oelhaf\textsuperscript{1}\textsuperscript{$\dagger$}\textsuperscript{*}, Georg Kordowich\textsuperscript{2}\textsuperscript{$\dagger$}, Christian Bergler\textsuperscript{3}, Andreas Maier\textsuperscript{1}, Johann J{\"a}ger\textsuperscript{2}, Siming Bayer\textsuperscript{1}
}

\IEEEauthorblockA{
    \textit{\textsuperscript{1}Pattern Recognition Lab, Friedrich-Alexander-Universität Erlangen-Nürnberg, Erlangen, Germany}\\
    \textit{\textsuperscript{2}Institute of Electrical Energy Systems, Friedrich-Alexander-Universität Erlangen-Nürnberg, Erlangen, Germany}\\
    \textit{\textsuperscript{3}Department of Electrical Engineering, Media and Computer Science,}\\
    \textit{Ostbayerische Technische Hochschule Amberg-Weiden, Amberg, Germany}\\
    {\footnotesize \textsuperscript{$\dagger$}These authors contributed equally.}\\
    {\footnotesize \textsuperscript{*}Corresponding author: julian.oelhaf@fau.de}
}
}

\begin{document}
\newacro{ai}[AI]{artificial intelligence}
\newacro{ml}[ML]{machine learning}
\newacro{dl}[DL]{deep learning}
\newacro{cnn}[CNN]{convolutional neural network}
\newacro{rnn}[RNN]{recurrent neural network}
\newacro{lstm}[LSTM]{long short-term memory}
\newacro{gru}[GRU]{gated recurrent unit}
\newacro{ann}[ANN]{artificial neural network}
\newacro{tft}[TFT]{temporal fusion transformer}
\newacro{tcn}[TCN]{temporal convolution network}

\newacro{mlp}[MLP]{multi-layer perceptron}
\newacro{knn}[KNN]{k-nearest neighbors}
\newacro{dt}[DT]{decision tree}

\newacro{dae}[DAE]{deep autoencoder}
\newacro{svm}[SVM]{support vector machine}
\newacro{dbn}[DBN]{deep belief network}
\newacro{tkeo}[TKEO]{Teager–Kaiser energy operator}
\newacro{wams}[WAMS]{wide-area measurement system}

\newacro{fc}[FC]{fault classification}
\newacro{fd}[FD]{fault detection}
\newacro{fl}[FL]{fault localization}
\newacro{fli}[FLI]{fault line identification}

\newacro{ibr}[IBR]{inverter-based resources}
\newacro{res}[RES]{renewable energy sources}
\newacro{der}[DER]{distributed energy resources}
\newacro{hvdc}[HVDC]{high-voltage direct current}
\newacro{mt-hvdc}[MT-HVDC]{multi-terminal high-voltage direct current}
\newacro{vsc}[VSC]{voltage source converter}

\newacro{pr}[PR]{protection relay}
\newacro{rtds}[RTDS]{real-time digital simulator}
\newacro{dwt}[DWT]{discrete wavelet transform}
\newacro{fft}[FFT]{fast Fourier transform}
\newacro{tsne}[t-SNE]{t-distributed stochastic neighbor embedding}
\newacro{pmu}[PMU]{phasor measurement unit}

\newacro{emt}[EMT]{electromagnetic transient}

\maketitle

\begin{abstract}
The increasing interest in data-driven methods for power system protection is accompanied by a lack of standardized, publicly available high-voltage waveform datasets that enable transparent and reproducible evaluation. To address this gap, this paper introduces the PROTECT-90 dataset, an open electromagnetic transient (EMT)-simulated reference benchmark for high-voltage fault studies with consistent digital-fault-recorder-like measurements, publicly released with this work. The dataset comprises 9{,}022 physically consistent short-circuit simulation episodes generated on a standardized 90~kV double-line topology with systematically documented domain randomization of grid operating points, line parameters, and fault conditions. For each episode, synchronized three-phase voltage and current waveforms are recorded at eight measurement locations and released together with structured, machine-readable metadata describing fault type, fault location, inception time, and operating conditions. All modeling assumptions, parameter ranges, and data-generation procedures are explicitly documented to ensure transparency and cross-study comparability. By combining physically grounded EMT simulation, balanced scenario coverage, and open accessibility, PROTECT-90 establishes a standardized foundation for reproducible benchmarking of protection-oriented signal processing and learning-based methods.
\end{abstract}


\section{Introduction}
\label{sec:introduction}

Reliable high-voltage-level protection relies on accurate interpretation of voltage and current waveforms during short-circuit events. \Ac{emt} simulation is widely used to analyze fault inception and fast transient phenomena not captured by steady-state phasor-domain models. As high-voltage systems operate under increasingly diverse conditions, including varying load levels, short-circuit strengths, voltage setpoints, and line characteristics, protection validation requires waveform data that reflect realistic operating variability. Despite the central role of \ac{emt} simulation in protection engineering, openly accessible waveform datasets with clearly documented topology definitions, parameter ranges, and labeling conventions remain scarce. Many studies rely on proprietary grid models or task-specific data generation procedures that are not publicly released, restricting reproducibility and cross-study comparability.

Standardized test systems such as the IEEE 9-bus and IEEE 39-bus networks are widely used for steady-state and phasor-domain validation~\cite{athay_practical_1979,vittal_power_2020}. For waveform-based analysis, publicly available resources include \ac{emt}-simulated datasets such as the Inverter-Rich Transmission System Disturbance (IRTSD) dataset~\cite{brett_ross_irtsd_2024}, as well as measured waveform repositories including the SoCal 28-Bus distribution grid dataset (SoCal)~\cite{xie_digital_2025}, the RTE digital fault recording database~\cite{presvots_database_2024}, and the Grid Event Signature Library (GESL)~\cite{wilson_grid_2024}. As summarized in Table~\ref{tab:dataset_comparison}, these resources provide either synthetic events tailored to specific grid models or real-world recordings without explicitly documented and controllable system parameter bounds, limiting their suitability for transparent and reproducible \ac{emt}-based benchmarking.

Taken together, existing public resources either provide controlled \ac{emt} simulations without releasing complete waveform corpora and parameter metadata, or offer measured recordings without structured and fully documented simulation assumptions. As reported in a recent scoping review of machine learning applications in power system protection~\cite{oelhaf_scoping_2025}, only 3.4\,\% of surveyed studies released their datasets or relied on publicly accessible data, with most based on proprietary simulations or unpublished grid models. Moreover, topology definitions, parameter ranges, and data generation procedures are often only partially documented, limiting independent reproduction of reported results. To the best of our knowledge, no openly available high-voltage level \ac{emt} waveform dataset simultaneously provides raw instantaneous measurements, complete machine-readable metadata, and explicitly documented parameter bounds for systematic and reproducible evaluation. This lack of transparent and standardized waveform corpora remains a structural barrier to cross-study comparability.

\newif\ifdebugtopology
\debugtopologyfalse     

\begin{figure*}[t]
    \centering
    \begin{tikzpicture}[transform shape]
        \node[anchor=south west, inner sep=0] (img) at (0,0)
            {\includegraphics[width=0.99\linewidth]{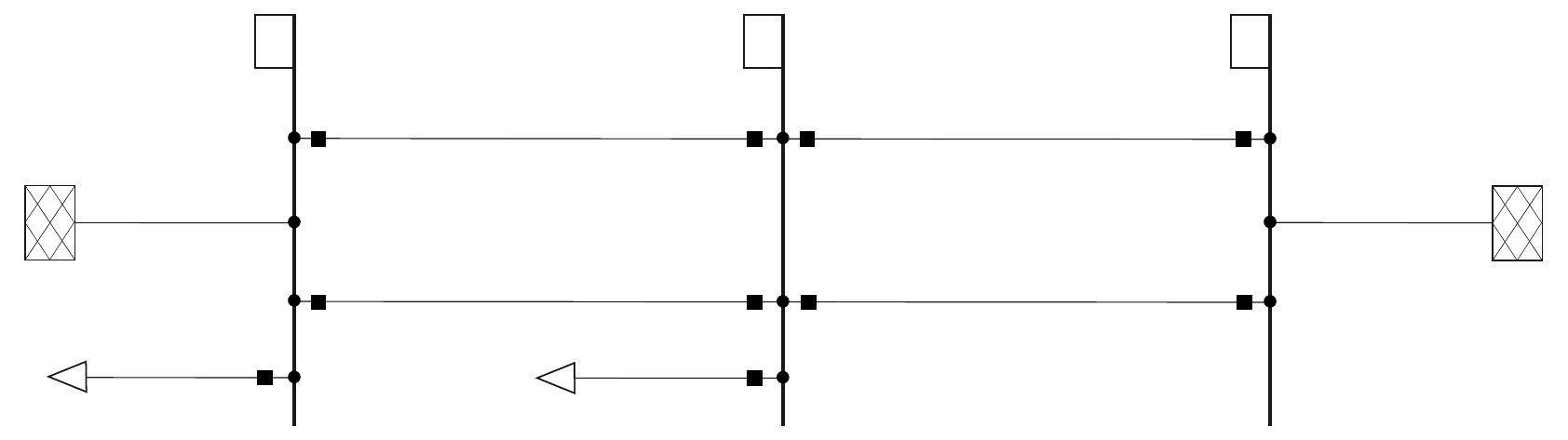}};
        \begin{scope}[x={(img.south east)}, y={(img.north west)}]

            \ifdebugtopology
            
            \draw[green!80!black, dashed, thick] (0.1725,0) -- (0.1725,1);
            \draw[green!80!black, dashed, thick] (0.484,0) -- (0.484,1);
            \draw[green!80!black, dashed, thick] (0.7945,0) -- (0.7945,1);

            \fi

            \node[inner sep=2pt, font=\small] at (0.32,0.73) {Line 2--3 A};
            \node[inner sep=2pt, font=\small] at (0.32,0.36) {Line 2--3 B};

            \node[inner sep=2pt, font=\small] at (0.63,0.73) {Line 1--2 A};
            \node[inner sep=2pt, font=\small] at (0.63,0.36) {Line 1--2 B};

            \node[inner sep=2pt, font=\bfseries] at (0.7945+0.05,0.96) {Bus 1};
            \node[inner sep=2pt, font=\bfseries] at (0.484+0.05,0.96) {Bus 2};
            \node[inner sep=2pt, font=\bfseries] at (0.1725+0.05,0.96) {Bus 3};

            \node[inner sep=2pt, font=\small] at (0.96,0.65) {Ext. Grid 1};
            \node[inner sep=2pt, font=\small] at (0.04,0.65) {Ext. Grid 3};

            \node[inner sep=2pt, font=\small] at (0.035,0.04) {Load 3};
            \node[inner sep=2pt, font=\small] at (0.35,0.04) {Load 2};

            \node[anchor=west, font=\scriptsize] at (0.00,-0.1)
            {\rule{5pt}{5pt}\, Protection relay location};

        \end{scope}
    \end{tikzpicture}

    \caption{Standardized 90~kV high-voltage double-line topology used for EMT dataset generation. Filled squares denote protection relay measurement locations.}
    \label{fig:double_line_grid_topology}
\end{figure*}
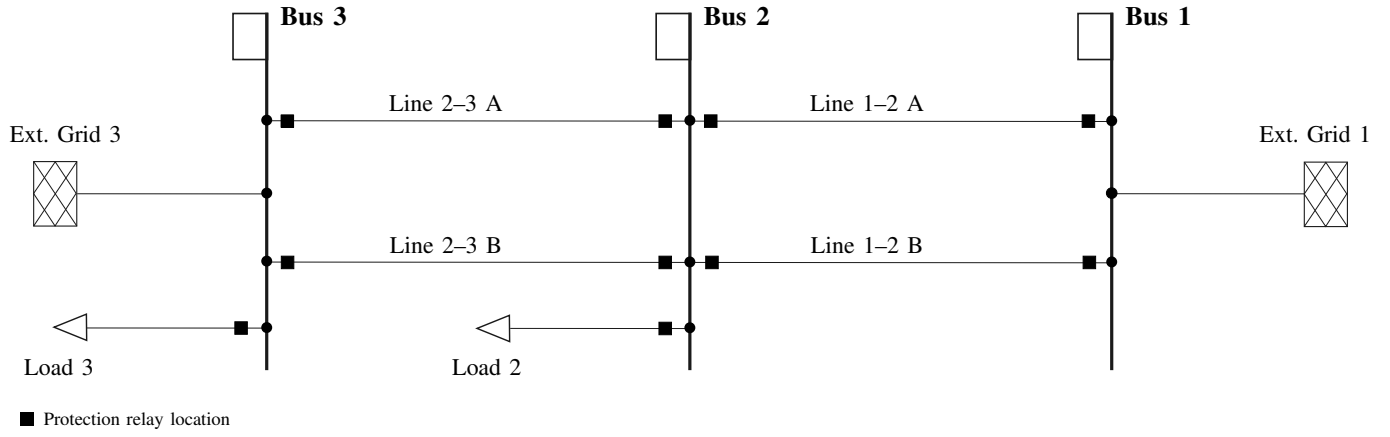

This paper addresses this gap by introducing PROTECT-90, a publicly released 90\,kV \ac{emt}-simulated high-voltage level waveform dataset comprising 9{,}022 physically consistent simulation episodes. The dataset provides raw instantaneous three-phase voltage and current measurements together with structured, machine-readable metadata describing fault characteristics, grid parameters, and topology states. A central design principle is the use of \emph{physically constrained domain randomization}, whereby operating conditions are systematically varied within explicitly defined and physically plausible bounds to enable controlled yet diverse scenario generation.

The main contributions of this work are:
\begin{itemize}
    \item The release of PROTECT-90, a publicly accessible high-voltage \ac{emt} waveform dataset with 9{,}022 labeled fault scenarios and structured metadata available at \textit{\href{https://doi.org/10.5281/zenodo.18418330}{10.5281/zenodo.18418330}}.
    \item A physically constrained domain-randomization framework with explicitly documented parameter bounds.
    \item Full documentation of modeling assumptions and data-generation procedures under an open license, enabling transparent and reproducible benchmarking.
\end{itemize}

\section{Design Rationale and Scope}

The PROTECT-90 dataset is designed as a standardized high-voltage \ac{emt} reference case rather than a large-scale meshed network model. A nominal voltage level of 90~kV is selected to represent a widely deployed sub-transmission class while maintaining alignment with practical high-voltage digital fault recording conditions. This choice enables realistic transient behavior without restricting the dataset to extra-high-voltage or inverter-dominated system configurations.

The adopted double-line topology reflects common high-voltage corridor structures featuring parallel current paths and intermediate infeeds~\cite{ziegler_numerical_2011}. These characteristics introduce non-trivial current distribution during short-circuit events while remaining sufficiently compact for controlled parameter variation and full documentation. The topology is intentionally fixed to provide a reproducible and interpretable reference scenario, enabling systematic comparison across studies.

The scope of the dataset is limited to  \ac{emt} simulations under physically constrained domain randomization of grid and fault parameters. It does not aim to represent large meshed networks or specific protection schemes but instead provides a transparent and standardized waveform corpus suitable for reproducible transient analysis and protection-oriented research.

\section{Simulation Framework}

This section describes the electrical modeling assumptions, \ac{emt} simulation configuration, and measurement structure used to generate the dataset.

\subsection{Benchmark Topology}

The benchmark topology, shown in Fig.~\ref{fig:double_line_grid_topology}, is a standardized 90~kV high-voltage double-line network comprising three buses (Bus~1--3) connected by two parallel corridors between adjacent bus pairs. Line~2--3~A/B connect Bus~2 and Bus~3, and Line~1--2~A/B connect Bus~1 and Bus~2, forming parallel current paths commonly encountered in practice~\cite{ziegler_numerical_2011}. Such configurations introduce non-trivial current distribution and impedance measurement errors relevant for distance protection studies.

Individual line lengths and electrical parameters are sampled within predefined bounds (Table~\ref{tab:emt_parameter_ranges}). The selected parameter ranges are chosen to represent realistic high-voltage conditions while covering a broad variability spectrum. The bounds are derived from established literature on short-circuit current calculation and high-voltage system modeling~\cite{roeper_kurzschlusstrome_1984,oeding_elektrische_2016}. Bus~1 and Bus~3 are connected to external grid equivalents, and aggregated loads are connected at Bus~2 and Bus~3, resulting in multiple power-flow paths during fault conditions.
To introduce structural variability, the secondary corridor segments (Line~1--2~B and Line~2--3~B) are switched out in a subset of scenarios, allowing different configurations and their respective impact on power system protection.

\subsection{EMT Simulation Configuration and Measurement Model}

All simulations are performed in \ac{emt} mode using DIgSILENT PowerFactory\footnote{\url{https://www.digsilent.de/en/powerfactory.html}}. For each scenario, the system is first simulated for an internal warm-up period of 0.1~s to allow the differential equation system to settle to steady-state conditions. The subsequent 1~s analysis window is then recorded and exported, sampled at 6{,}400\,Hz, resulting in 6{,}400 time steps per episode. The 6.4\,kHz sampling rate corresponds to 128 samples per 50\,Hz fundamental cycle, representing a typical digital fault recording resolution in high-voltage protection while enabling high-resolution transient analysis with manageable data size.
A fixed-step \ac{emt} solver with a step size of 10~$\mu$s is used to ensure consistent and accurate results across all simulations. Three-phase voltages and currents are recorded at both terminals of each line segment, yielding six measurement channels per relay location and a total of 48 synchronized waveform channels. To ensure consistent signal availability across all scenarios, voltage and current signals are extracted directly at the corresponding bus terminals within the simulation model. No secondary instrument transformer models (CTs/VTs) are considered; all exported quantities represent ideal primary voltage and current signals.

\begin{table}[t]
\centering
\begin{threeparttable}
\caption{Positioning of publicly available waveform datasets (IRTSD~\cite{brett_ross_irtsd_2024}, SoCal~\cite{xie_digital_2025}, RTE~\cite{presvots_database_2024}, GESL~\cite{wilson_grid_2024}).}
\label{tab:dataset_comparison}
\renewcommand{\arraystretch}{1.05}
\setlength{\tabcolsep}{4pt}
\begin{tabular}{p{2.4cm}ccccc}
\toprule
\textbf{Feature} & \textbf{PROTECT-90} & \textbf{IRTSD} & \textbf{SoCal} & \textbf{RTE} & \textbf{GESL} \\
\midrule
EMT-simulated & $\checkmark$ & $\checkmark$ & -- & -- & -- \\
Measured waveforms & -- & -- & $\checkmark$ & $\checkmark$ & $\checkmark$ \\
High-voltage focus & $\checkmark$ & $\checkmark$ & -- & $\checkmark$ & -- \\
Explicit parameter bounds documented & $\checkmark$ & -- & -- & -- & -- \\
\bottomrule
\end{tabular}

\begin{tablenotes}
\footnotesize
\item Notes: SoCal, RTE, and GESL provide real-world waveform measurements, while IRTSD and the proposed dataset are EMT-simulated. The proposed dataset is designed as a high-voltage level benchmark with explicitly documented parameter bounds.
\end{tablenotes}

\end{threeparttable}
\end{table}

\section{Scenario Generation and Automation}

The automation concept underlying the scenario generation process extends a previously published generic data generation framework for short-circuit studies in PowerFactory~\cite{wang_generic_2022}. While the present implementation was redesigned and developed independently for the structured PROTECT-90 benchmark, the earlier work established the principle of automated fault injection, parameter randomization, and parallelized simulation execution.
Each simulation scenario (episode) contains a single short-circuit event. Faults are defined by the affected line segment, fault type (SLG, LL, LLG, LLL), and involved phase(s). The fault resistance $R_f$ and inception time $t_f$ are selected within the ranges specified in Table~\ref{tab:emt_parameter_ranges}. Fault locations are uniformly distributed along the normalized line length, ensuring spatial coverage across the full extent $\ell$.

To represent operating variability, grid parameters are varied according to a domain randomization strategy. The objective is to generate a broad spectrum of physically plausible operating conditions, thereby increasing waveform diversity and improving the robustness of data-driven methods with respect to domain shift from simulation to real-world measurements. The electrical parameters listed in Table~\ref{tab:emt_parameter_ranges} are independently sampled within their specified bounds prior to each simulation. These include the line length $\ell$, series resistance $R'$, series reactance $X'$, and shunt capacitance $C'$ per kilometer, as well as the load active and reactive power $(P,Q)$.

External grid parameters -- short-circuit power $S_k''$, voltage magnitude $V$, and voltage angle $\phi$ -- are likewise varied within their respective bounds. The selected short-circuit power range is intentionally biased toward comparatively low values to reflect future grid conditions with increasing penetration of inverter-based resources and consequently reduced fault levels. The voltage angle $\phi$ is randomized to prevent systematic phase alignment across scenarios and avoid artificial waveform synchronization. In addition, the external grid at Bus~3 is deactivated in a subset of cases to emulate reduced infeed or topology reconfiguration.

To ensure physical realism, all parameter combinations must satisfy load-flow convergence and \ac{emt} numerical stability before, during, and after the fault event. Furthermore, the $R/X$ ratio of all lines is constrained to the interval $[0.05, 0.5]$, reflecting typical high-voltage line characteristics. This physically constrained randomization preserves electrical plausibility while maintaining a wide variability spectrum representative of real-world high-voltage system operation.

\begin{table}[t]
\centering
\caption{Parameter ranges used for EMT domain randomization.}
\label{tab:emt_parameter_ranges}
\setlength{\tabcolsep}{6pt}
\renewcommand{\arraystretch}{1.03}
\begin{tabular}{p{3.4cm} c c c}
\toprule
\textbf{Parameter} & \textbf{Unit} & \textbf{Min} & \textbf{Max} \\
\midrule
\multicolumn{4}{l}{\emph{Fault Parameters}} \\
Fault resistance $R_f$ & $\Omega$ & 0.1 & 10 \\
Fault inception time $t_f$ & s & 0.2 & 0.5 \\
\midrule
\multicolumn{4}{l}{\emph{Line Parameters}} \\
Line length $\ell$ & km & 10 & 60 \\
Series resistance per km $R'$ & $\Omega$/km & 0.01 & 0.20 \\
Series reactance per km $X'$ & $\Omega$/km & 0.35 & 0.45 \\
Shunt capacitance per km $C'$ & nF/km & 8.5 & 10 \\
\midrule
\multicolumn{4}{l}{\emph{Load Parameters}} \\
Active power $P$ & MW & 20 & 50 \\
Reactive power $Q$ & Mvar & $-20$ & 20 \\
\midrule
\multicolumn{4}{l}{\emph{External Grid Parameters}} \\
Short-circuit power $S_k''$ & MVA & 90 & 1000 \\
Voltage magnitude $V$ & pu & 0.95 & 1.05 \\
Voltage angle $\phi$ & deg & $-180$ & 180 \\
\bottomrule
\end{tabular}
\end{table}

\section{PROTECT-90: Dataset Structure and Coverage}\label{sec:dataset_structure}

Each simulation episode is stored as an individual \texttt{.pkl} file containing a time-series dataframe with 49 columns. The first column corresponds to the time vector, while the remaining 48 columns represent synchronized three-phase voltage and current measurements from eight relay locations (six channels per relay).

Signal naming follows a consistent convention combining bus, line section, quantity, phase, and unit, e.g.,
\texttt{Bus\_3\_Line\_02\_03A\_vol\_L3\_V}, denoting the phase~3 voltage at the Bus~3 terminal of line section~2--3A. Currents are labeled analogously using the suffix \texttt{\_cur\_Lx\_A}. All quantities correspond to secondary voltage (V) and current (A) signals sampled at 6{,}400\,Hz.

\begin{figure}[t]
    \centering
    \includegraphics[width=\linewidth]{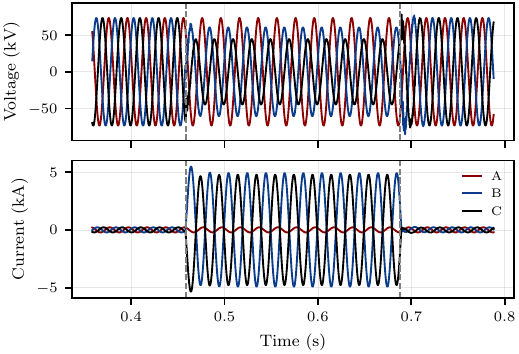}
    \caption{Representative three-phase voltage (top) and current (bottom) waveforms for a two-phase short-circuit between phases B and C on Line~1--2~B. The fault is applied at 31.8\,\% of the line length with a resistance of 6.56~$\Omega$. Dashed vertical lines indicate fault inception ($t = 0.458$\,s) and the end of the injected fault (fault element removed at $t = 0.688$\,s).}
    \label{fig:waveform_example}
\end{figure}

At the episode level, the dataset can be formally represented as
\begin{equation*}
X \in \mathbb{R}^{N_{\mathrm{PR}} \times 6 \times T_s},
\end{equation*}
where $N_{\mathrm{PR}}=8$ denotes the number of relay locations, six channels correspond to three-phase voltage and current measurements per relay, and $T_s=6{,}400$ is the number of time steps per episode. A representative example of the recorded three-phase voltage and current waveforms during a fault event is shown in Figure~\ref{fig:waveform_example}.

Structured metadata for all episodes are provided in a CSV file indexed by a unique \texttt{sample\_id}. Metadata fields include fault type, affected phases, faulted line section, normalized fault location, fault resistance, inception time, randomized grid parameters (with physical units), and topology switching states (boolean flags). No derived or preprocessed features are included, allowing users to define task formulations explicitly. A concise summary of the dataset characteristics and core technical specifications is provided in Table~\ref{tab:dataset_overview}, facilitating transparent reporting and cross-study comparability.

\begin{table}[t]
\centering
\caption{Overview and key characteristics of the PROTECT-90 dataset.}
\label{tab:dataset_overview}
\renewcommand{\arraystretch}{1.06}
\setlength{\tabcolsep}{4pt}
\begin{tabular}{p{2.4cm} p{5.6cm}}
\toprule
\textbf{Property} & \textbf{Specification} \\
\midrule
\textbf{Topology} & 90\,kV double-line system \\
\textbf{Simulation type} & Electromagnetic transient (EMT) \\
\textbf{Sampling rate} & 6.4\,kHz (128 samples per 50\,Hz cycle) \\
\textbf{Episode duration} & 1\,s (6{,}400 samples) \\
\textbf{Episodes} & 9{,}022 fault scenarios \\
\textbf{Location} & Normalized line length \\
\textbf{Domain Randomization} & Line, load, grid, and fault parameters (Table~\ref{tab:emt_parameter_ranges}) \\
\textbf{Metadata} & CSV (fault and grid parameters) \\
\textbf{Train/test split} & None (episode-wise partitioning recommended) \\
\textbf{License} & CC BY 4.0 \\
\textbf{DOI} & \href{https://doi.org/10.5281/zenodo.18418330}{10.5281/zenodo.18418330} \\
\bottomrule
\end{tabular}
\end{table}

\subsection{Coverage and Dataset Characteristics}

\begin{figure*}[t]
  \centering
  \includegraphics[width=\linewidth]{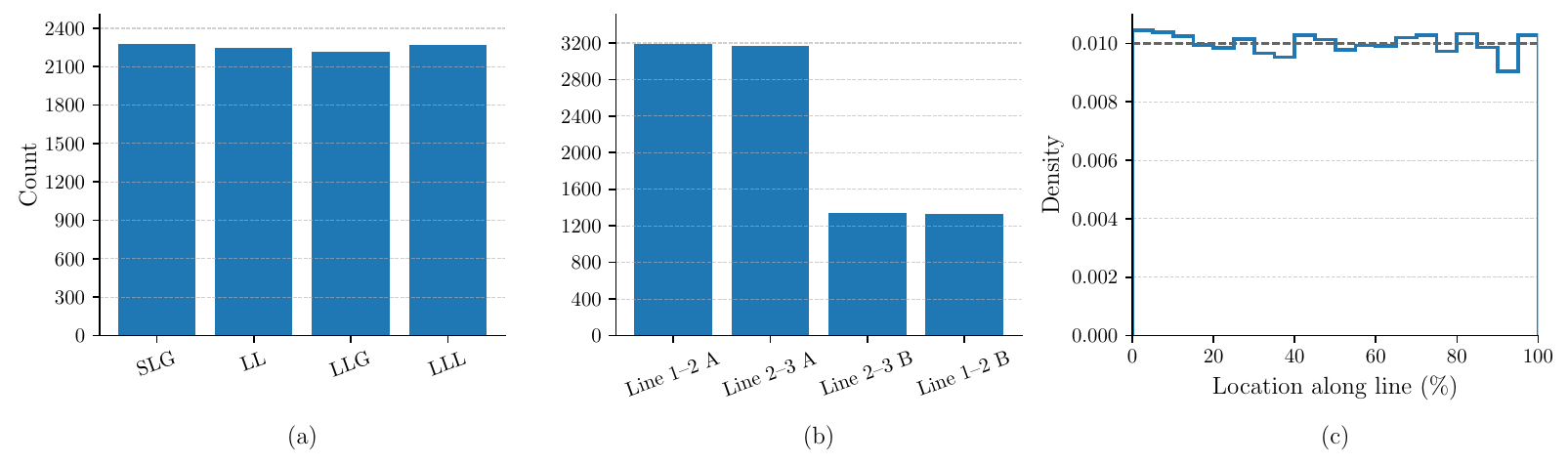}
  \caption{Coverage statistics of the \ac{emt} dataset. (a) Distribution of short-circuit fault types, showing balanced sampling across SLG, LL, LLG, and LLL categories. (b) Distribution of faulted line segments, reflecting corridor switching states and resulting episode counts per segment. (c) Spatial fault-location distribution along the line segments, illustrating approximately uniform sampling across the full line length.}
  \label{fig:coverage_stats}
\end{figure*}

The dataset comprises 9{,}022 fault scenarios without a non-fault class. Figure~\ref{fig:coverage_stats} summarizes the distribution of fault types, faulted line sections, and spatial fault locations. Fault categories (SLG, LL, LLG, LLL) are approximately uniformly distributed within each line section. The primary corridor sections (1--2A and 2--3A) each account for roughly 35\,\% of the dataset, while the secondary corridor sections (1--2B and 2--3B) contribute approximately 15\,\% each, reflecting the applied topology switching strategy. The spatial distribution of fault locations is approximately uniform along the full line length.

Topology switching of Line~1--2B, Line~2--3B, and the external grid at Bus~3 is balanced (approximately 50/50), ensuring representation of both single- and double-circuit configurations as well as reduced infeed conditions.

Basic integrity checks are performed prior to export. Simulations that fail load-flow convergence are discarded. Exported episodes are verified to contain no missing values. No predefined train/test split is provided; dataset partitioning is intentionally left to the user to enable application-specific evaluation protocols.

Overall, the waveform structure, metadata schema, and balanced scenario coverage establish a reproducible and systematically documented benchmark for high-voltage fault studies.

\section{Reproducibility and Data Access}

PROTECT-90 is publicly available via Zenodo under DOI: \textit{\href{https://doi.org/10.5281/zenodo.18418330}{10.5281/zenodo.18418330}}~\cite{kordowich_protect-90_2026} and released under the Creative Commons Attribution 4.0 International (CC BY 4.0) license. The archive (12.5\,GB compressed, approximately 31\,GB uncompressed) contains the complete waveform corpus and all domain-randomized grid and fault parameters described in Table~\ref{tab:emt_parameter_ranges}. Waveforms are provided as Python pickle files (Pandas DataFrames, Python 3.11, \texttt{float32}), accompanied by a structured CSV metadata file (7.3\,MB) containing all labels and scenario parameters. No predefined train/test split is imposed, allowing users to define task-specific partitioning protocols.

\section{Intended Usage and Practical Recommendations}

The PROTECT-90 dataset and associated metadata enable multiple protection-oriented task formulations. For \emph{fault classification}~\cite{oelhaf_controlled_2025}, the fields \texttt{sc\_type} and \texttt{phase\_select} provide categorical targets. For \emph{fault line identification}~\cite{oelhaf_systematic_2025}, \texttt{fault\_target} serves as a discrete label. For \emph{fault localization}, the \texttt{sc\_location} variable enables regression-based estimation of the fault position. Randomized grid parameters (e.g., $S_k''$, $V$, $\ell$, $R'$, $X'$) allow controlled analysis of robustness and domain shift effects under varying operating conditions~\cite{oelhaf_impact_2025,oelhaf_robustness_2026}. Boolean topology flags support evaluation under single- and double-circuit configurations or reduced infeed conditions.

For benchmarking, episode-wise data partitioning is recommended. Sliding windows derived from the same simulation episode should not be distributed across training and test sets to prevent temporal leakage. Because the dataset contains only fault scenarios, detection tasks require explicit definition of windowing and labeling assumptions. Preprocessing steps, normalization procedures, and evaluation metrics should be reported to ensure comparable results across studies. The dataset is intentionally task-agnostic and does not prescribe specific window lengths, model architectures, or validation protocols, enabling flexible and application-dependent evaluation designs.

\section{Conclusion}

This paper presents PROTECT-90, a publicly released electromagnetic transient dataset for high-voltage fault studies based on a standardized 90~kV double-line topology. The dataset comprises 9{,}022 physically consistent simulation episodes with synchronized three-phase voltage and current measurements and fully documented domain-randomized grid and fault parameters. All modeling assumptions, parameter ranges, and topology configurations are explicitly disclosed, enabling transparent and reproducible research.

Unlike proprietary or partially documented waveform repositories, the dataset provides machine-readable metadata and complete traceability between measurements and underlying system conditions. By combining high-resolution EMT waveforms with structured parameter documentation, it establishes a physically grounded and fully documented benchmark for principled evaluation of protection-oriented signal processing and learning-based methods.

The dataset is openly accessible under a permissive license and intended to support reproducible benchmarking, cross-study comparison, and future extensions toward more complex grid topologies and measurement assumptions.

\section*{Acknowledgment}
This project was funded by the Deutsche Forschungsgemeinschaft (DFG, German Research Foundation) - 535389056.


\bibliographystyle{IEEEtran}
\bibliography{references}

\end{document}